\documentclass[12pt]{article}
\usepackage{epsfig,amssymb,amsmath,psfrag,subfigure,rotate,color,wasysym}

\allowdisplaybreaks
\newcommand{\insertfig}[2]{\includegraphics[width=#1cm]{#2}}

\def\XXint#1#2#3{{\setbox0=\hbox{$#1{#2#3}{\int}$ }
\vcenter{\hbox{$#2#3$ }}\kern-.6\wd0}}

\def \be  {\begin{equation}}
\def \ee  {\end{equation}}
\def \ba  {\begin{eqnarray}}
\def \ea  {\end{eqnarray}}
\def \baa {\begin{eqnarray*}}
\def \eaa {\end{eqnarray*}}
\def \lab #1 {\label{#1}}

\newcommand\re[1]{(\ref{#1})}
\def\d{\hbox{{d}\kern-.20em\hbox{l}}}

\def \matrix #1 {\left(\begin{array}{cc} #1 \end{array}\right)}

\newcommand \vev [1] {\langle{#1}\rangle}

\newcommand \ket [1] {|{#1}\rangle}
\newcommand \bra [1] {\langle {#1}|}
\newcommand{\bit}[1]{\mbox{\boldmath$#1$}}

\newcommand{\ft}[2]{{\textstyle\frac{#1}{#2}}}
\begin{document}

\begin{titlepage}

\thispagestyle{empty}

\vspace*{3cm}

\centerline{\large \bf Multichannel conformal blocks for scattering amplitudes}
\vspace*{1cm}

\centerline{\sc A.V.~Belitsky}

\vspace{10mm}

\centerline{\it Department of Physics, Arizona State University}
\centerline{\it Tempe, AZ 85287-1504, USA}

\vspace{2cm}

\centerline{\bf Abstract}

\vspace{5mm}

By performing resummation of small fermion-antifermion pairs within the pentagon form factor program to scattering amplitudes in planar $\mathcal{N} = 4$ 
superYang-Mills theory, we construct multichannel conformal blocks within the flux-tube picture for $N$-sided NMHV polygons. This procedure is equivalent to 
summation of descendants of conformal primaries in the OPE framework. The resulting conformal partial waves are determined by multivariable hypergeometric 
series of Lauricella-Saran type.

\end{titlepage}

\setcounter{footnote} 0

\newpage

%\pagestyle{plain}
%\setcounter{page} 1

%{
%\footnotesize 
%\tableofcontents}

%\newpage

{\bf 1. Introduction.} Symmetries of a system allow one to significantly reduce the number of degrees of freedom that require dynamical considerations. Conformal block decomposition 
of correlation functions $\vev{\prod_j \mathcal{O}_j}$ of local operators $\mathcal{O}_j \equiv \mathcal{O}_j (z_j)$ is a way of implementing them in a scale-invariant field theory (or 
CFT) via the operator product expansion (OPE). Under the assumption of convergence, a correlator can be expanded in a complete set of primary operators ${\Phi}_{\Delta_\ell}$ of 
increasing scaling dimension and spin (cumulatively called $\Delta_\ell$) and their conformal descendants built with the action of derivatives $\partial^n {\Phi}_\ell$. It is the latter infinite 
tower which is conveniently packed together in the conformal block, also known as the partial wave $ \mathcal{F}_{\bit{\scriptstyle\Delta}} (\bit{w})$, which is a function of $\bit{\Delta} 
= \{ \Delta_\ell \}$ and cross ratios $\bit{w} = \{ w_\ell \}$, schematically,
\begin{align}
\vev{\prod_j \mathcal{O}_j}
=
\left(
\prod_{j<k} z_{jk}^{\Delta_{jk}} 
\right)
\sum_{\bit{\scriptstyle\Delta}} a_{\bit{\scriptstyle\Delta}} \mathcal{F}_{\bit{\scriptstyle\Delta}} (\bit{w})
\, ,
\end{align}
with an overall multiplicative function of the coordinate differences with powers $\Delta_{jk}$ being functions of the operator $ \mathcal{O}_j$ dimensions/spins conveniently 
chosen to carry the scaling dimension of the left-hand side. The conformal blocks $\mathcal{F}_{\bit{\scriptstyle\Delta}}$ are eigenfunctions of conformal Casimir operators 
for successive channels in the operator product expansion and are subject to appropriate boundary conditions. While the low-point correlators are well studied, there is little
to no knowledge of multichannel conformal blocks.

Conformal blocks are ubiquitous in physics so they make their natural appearance in the analysis of scattering amplitudes within the pentagon operator product expansion
\cite{Basso:2013vsa,Alday:2010ku}. In the latter, one relies on a dual description of amplitudes in terms of excitations propagating on a color flux-tube sourced by the contour 
of the Wilson loop living in the four-dimensional momentum space \cite{Alday:2007hr,Drummond:2007cf,Brandhuber:2007yx,CaronHuot:2010ek,Mason:2010yk,Belitsky:2011zm}. 
The vacuum represented by the flux is in fact SL(2) invariant to lowest order in 't Hooft coupling \cite{Gaiotto:2010fk,Basso:2010in}. This property was used in the construction of 
conformal blocks for (N)MHV hexagons and heptagons \cite{Gaiotto:2011dt,Cordova:2016woh,Sever:2011pc,Belitsky:2017wdo}. 

The tree-level $N$-particle ratio function of the NMHV to MHV tree amplitudes 
\begin{align}
\label{RinvariantRatio}
\mathbb{R}_N = \sum_{1 < j < k < N - 2} [1, j, j + 1, k, k + 2]
\end{align}
is determined by the R-invariants \cite{Drummond:2008vq,Mason:2009qx}
\begin{align}
[i,j,k,l,m] = \frac{\delta^{0|4} (\chi_i (jklm) + \mbox{cyclic})}{(ijkl)(jklm)(klmi)(lmij)(mijk)}
\, , 
\end{align}
with the four-bracket defined by the determinant $(jklm) \equiv \varepsilon_{JKLM} Z_j^J Z_k^K Z_l^L Z_m^M$ built from the momentum twistors $Z_j^J$ and
$\chi_j^A$ being their fermionic partners. Within the pentagon form factor program, each individual Grassmann component $R^{[r_1, r_2, \dots, r_{N-5}]}$ of 
$\mathbb{R}_N$, with R-weights $r_1, \dots, r_{N-5}$ of all parent excitations, corresponding to the SU(4) dimensions ${\bf R}$ of flux-tube excitations,
$r = 0,1,2,3,4$ for ${\bf R} = \bar{\bf 1}, \bar{\bf 4}, {\bf 6}, {\bf 4}, {\bf 1}$, can be represented in terms of flux-tube integrals
\begin{align}
\label{NMHVcomponents}
R^{[r_1, r_2, \dots, r_{N-5}]} = \sum_{\alpha_1, \dots, \alpha_{N-5}} 
&
{\rm e}^{ - t_{\alpha_1} \tau_{1} - {\dots}  - t_{N-5} \tau_{\alpha_{N-5}} + i h_{\alpha_1} \varphi_1 + {\dots} + i h_{\alpha_{N-5}} \varphi_{N-5}}
\\
&\times
\int \prod_{j = 1}^{N-5} \frac{d u_j}{2 \pi} {\rm e}^{2 i \sigma_1 u_1 + \dots 2 i \sigma_{N-5} u_{N-5}} I^{{\bf\scriptscriptstyle R}_1 | \dots | {\bf\scriptscriptstyle R}_{N-5}}
(\alpha_1, u_1 | \dots | \alpha_{N-5}, u_{N-5})
\nonumber
\, ,
\end{align}
where the $3(N-5)$ conformal invariants of Eq.\ \re{RinvariantRatio} were traded for $N-5$ sets of triplets $(\tau_j, \sigma_j, \varphi_j)$ with their reciprocal variables 
interpreted as the energy (or twist), momentum and helicity, respectively, of the particles propagating on the flux and their SU(4) representation ${\bf R}_j$.

There is an infinite number of (parent) flux-tube excitations $\Phi^{\bf\scriptscriptstyle R}_\alpha$ \cite{Cordova:2016woh,Belitsky:2017wdo} of different spin/R-change 
and increasing energy (i.e., conformal primary states, in the language of CFT) which determine the integrand $I^{{\bf\scriptscriptstyle R}_1 | \dots | 
{\bf\scriptscriptstyle R}_{N-5}}$. Their descendants arise by gluing small fermion-antifermion pairs to $\Phi^{\bf\scriptscriptstyle R}_\alpha$'s. A small fermion-antifermion 
pair $\psi_{\rm s} \bar\psi_{\rm s}$ is equivalent to the derivative since $\psi_{\rm s}$ at zero momentum becomes the generator of Poincar\'e supersymmetry $Q$ and 
since $\{ Q, \bar{Q} \} \sim P$, according to their algebra, $(\psi_{\rm s} \bar\psi_{\rm s})^n \Phi^{\bf\scriptscriptstyle R}_\alpha \sim \partial^n \Phi^{\bf\scriptscriptstyle R}_\alpha$ 
by analogy with conformal OPE alluded to above. In this note, we will construct multichannel conformal blocks for $N$-leg NMHV amplitudes by explicit resummation of the 
entire tower of small fermion-antifermions pairs accompanying parent particles, this will yield the substitution in the integrand
\begin{align}
&
I^{{\bf\scriptscriptstyle R}_1 | \dots | {\bf\scriptscriptstyle R}_{N-5}} (\alpha_1, u_1 | \dots | \alpha_{N-5}, u_{N-5})
\\
&
\to 
I^{{\bf\scriptscriptstyle R}_1 | \dots | {\bf\scriptscriptstyle R}_{N-5}} (\alpha_1, u_1 | \dots | \alpha_{N-5}, u_{N-5})
\mathcal{F}^{[r_1, \dots, r_{N-5}]}_{h_{\alpha_1}, t_{\alpha_1} | \dots | h_{\alpha_{N-5}}, t_{\alpha_{N-5}}}  (u_1, \tau_1 | \dots | u_{N-5}, \tau_{N-5})
\nonumber
\end{align}
where $\mathcal{F}$ are the conformal blocks in question. This formalism is equivalent to the projection technique for computation of conventional conformal blocks in a 
CFT, which we briefly review by applying it to a four-point correlator in Appendix A to draw a parallel with the flux-tube physics.

%%%%%%%%%%%%%%%%%%%%%%%%%%%%%%%%%%%%%%%%%%%%%%%%%%%%%%%%%%%%%%%%%%%%%
%            Figure
%%%%%%%%%%%%%%%%%%%%%%%%%%%%%%%%%%%%%%%%%%%%%%%%%%%%%%%%%%%%%%%%%%%%%
\begin{figure}[t]
\begin{center}
\mbox{
\begin{picture}(0,260)(140,0)
\put(0,-80){\insertfig{16}{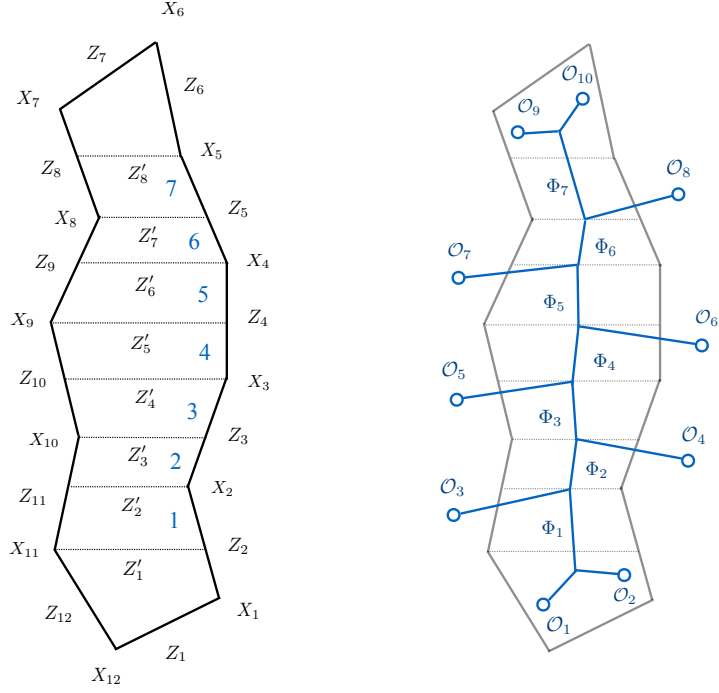}}
\end{picture}
}
\end{center}
\caption{ \label{DodecagonPic} A tessellation of a polygon (dodecagon on the left) and its OPE dual graph (on the right).
}
\end{figure}
%%%%%%%%%%%%%%%%%%%%%%%%%%%%%%%%%%%%%%%%%%%%%%%%%%%%%%%%%%%%%%%%%%%%%

{\bf 2. Kinematics.} Before turning to dynamics, let us introduce some kinematics first. The starting point is a tessellation of a polygon determined by the reference 
momentum twistors $Z_j$ in terms of a sequence of squares formed by the polygon edges and internal light-like lines encoded in the momentum twistors $Z'_k$  
(see the left panel in Fig.\ \ref{DodecagonPic} for the case of the dodecagon). A choice of a square automatically defines a conformal frame and thus a channel for 
propagation of parent flux excitations and their descendants. This is equivalent to a choice of an OPE channel for correlation functions (see the right panel in Fig.\ 
\ref{DodecagonPic}). To make the discussion more explicit, let us provide a choice of reference twistors for the dodecagon as a case of study (shown in Fig.\ \ref{DodecagonPic})
\begin{align}
\begin{array}{llll}
Z_1 = (6, 4, 12, 5)
\, , \quad
&
Z_2 = (1, 2, 4, 1)
\, , \quad
&
Z_3 = (0, 1, 1, 0)
\, , \quad
&
Z_4 = (0, 1, 0, 0)
\, , \\
Z_5 = (0, 2, -1, 1)
\, , \quad
&
Z_6 = (-1, 6, -4, 6)
\, , \quad
&
Z_7 = (-4, 6, -5, 12)
\, , \quad
&
Z_8 = (-2, 1, -1, 4)
\, , \\
Z_9 = (-1, 0, 0, 1)
\, , \quad
&
Z_{10} = (1, 0, 0, 0)
\, , \quad
&
Z_{11} = (2, 0, 1, 1)
\, , \quad
&
Z_{12} = (6, 1, 6, 4)
\, ,
\end{array}
\end{align}
while the twistors connecting the cusps $X_j$ with opposite sites of the polygon are
\begin{align}
\begin{array}{llll}
Z'_1 = (4, 1, 5, 3)
\, , \quad
&
Z'_2 = (1, 1, 3, 1)
\, , \quad
&
Z'_3 = (1, 0, 1, 1)
\, , \quad
&
Z'_4 =  (0 ,0, 1, 0)
\, , \\
Z'_5 = (0, 0, 0, 1)
\, , \quad
&
Z'_6 = (0, 1, -1, 1)
\, , \quad
&
Z'_7 = (-1, 1, -1, 3)
\, , \quad
&
Z'_8 = (-1, 4, -3, 5)
\, .
\end{array}
\end{align}

Every intermediate square enjoys a residual three-parameter conformal symmetry which leaves it invariant. These three parameters corresponds to the triplet 
$(\tau_j, \sigma_j, \varphi_j)$ introduced above. The invariance matrices for the squares can be determined successively starting with the middle one, i.e., fourth 
square in Fig. \ref{DodecagonPic}, which reads
\begin{align}
M_4  (\tau, \sigma, \phi) = {\rm diag} \left( {\rm e}^{\sigma - i \phi/2} \, ,  {\rm e}^{-\sigma - i \phi/2} \, , {\rm e}^{\tau + i \phi/2} \, , {\rm e}^{- \tau + i \phi/2} \right)
\, ,
\end{align}
and its matrix elements fixed in a particular conformal frame as recalled in Appendix B. The symmetry transformations for the rest can be obtained by finding rotation 
matrices of the corresponding twistors defining adjacent squares\footnote{For instance, the twistors for the 5-th square $\mathbb{Z}_5 = (Z'_6, Z_4, Z'_5, Z_9)$ can 
be determined from the 4-th one $\mathbb{Z}_4 = (Z'_4, Z_4, Z'_5, Z_{10})$ via the transformation $\mathbb{Z}_5 = R_5 \mathbb{Z}_4$.} and then using them 
for construction of the $M$-matrices, namely,
\begin{align}
M_7 (\tau_7, \sigma_7, \phi_7) &= R_6^{-1} M_6  (\tau_7, \sigma_7, \phi_7) R_6
\, , \nonumber\\
M_6 (\tau_6, \sigma_6, \phi_6) &= R_5^{-1} M_5  (\tau_6, \sigma_6, \phi_6) R_5
\, , \nonumber\\
M_5 (\tau_5, \sigma_5, \phi_5) &= R_4^{-1} M_4 (\tau_5, \sigma_5, \phi_5) R_4
\, , \nonumber\\
M_3 (\tau_3, \sigma_3, \phi_3) &= R_3^{-1} M_4  (\tau_3, \sigma_3, \phi_3) R_3
\, , \nonumber\\
M_2 (\tau_2, \sigma_2, \phi_2) &= R_2^{-1} M_3  (\tau_2, \sigma_2, \phi_2) R_2
\, , \nonumber\\
M_1 (\tau_1, \sigma_1, \phi_1) &= R_1^{-1} M_2  (\tau_1, \sigma_1, \phi_1) R_1
\, , 
\end{align}
where 
\begin{align}
\label{DodecagonTwistors}
\begin{array}{lll}
R_1 
= 
\left(
{\scriptsize
\begin{array}{cccc}
 10 & 2 & 11 & 7 \\
 -4 & 4 & 4 & -2 \\
 5 & -2 & 0 & 3 \\
 -23 & -2 & -21 & -16 \\
\end{array}
}
\right)
\, , 
&
R_2
=
\left(
{\scriptsize
\begin{array}{cccc}
 0 & 1 & 1 & 0 \\
 1 & 0 & 0 & 0 \\
 1 & 0 & 1 & 1 \\
 0 & 0 & 1 & 0 \\
\end{array}
}
\right)
\, ,
& 
R_3
=
\left(
{\scriptsize
\begin{array}{cccc}
 0 & 1 & 1 & 0 \\
 1 & 0 & 0 & 0 \\
 1 & 0 & 1 & 1 \\
 0 & 0 & 1 & 0 \\
\end{array}
}
\right)
\, , \\
& & \\
R_4
=
\left(
{\scriptsize
\begin{array}{cccc}
 0 & 1 & 0 & 0 \\
 1 & 0 & 0 & -1 \\
 0 & 0 & 0 & 1 \\
 0 & -1 & 1 & -1 \\
\end{array}
}
\right)
\, , 
&
R_5
=
\left(
{\scriptsize
\begin{array}{cccc}
 0 & 1 & -1 & 2 \\
 0 & -2 & 1 & -1 \\
 -1 & -2 & 1 & 1 \\
 -1 & 1 & -1 & 3 \\
\end{array}
}
\right)
\, , 
&
R_6 
=
\left(
{\scriptsize
\begin{array}{cccc}
 16 & -16 & 14 & -40 \\
 2 & 0 & 2 & -6 \\
 14 & -21 & 19 & -44 \\
 10 & -13 & 11 & -28 \\
\end{array}
}
\right)
\, .
\end{array}
\end{align}
In order to generate all inequivalent polygons, we act with these transformations on the twistors located either above or below it. For the case at hand, we have
\begin{align}
Z_1 
&
\to Z_1 M_1 M_2 M_3 \, , \qquad\quad
Z_2 \to Z_2 M_2 M_3 \, , \quad
Z_3 \to Z_3  \, , \quad \ 
Z_4 \to Z_4  
\, , \\
Z_5 
&
\to Z_5 M_5^{-1} M_4^{-1} \, , \qquad\quad
Z_6 \to Z_6 M_7^{-1} M_6^{-1} M_5^{-1} M_4^{-1} \, , \quad\
Z_7 \to Z_7 M_7^{-1} M_6^{-1} M_5^{-1} M_4^{-1} 
\, , \nonumber\\
Z_8 
&
\to Z_8 M_6^{-1} M_5^{-1} M_4^{-1} \, , \quad
Z_9 \to Z_9 M_4^{-1} \, , \quad
Z_{10} \to Z_{10} \, , \quad \ 
Z_{11} \to Z_{11}  M_3 \, , \quad
Z_{12} \to Z_{12} M_1 M_2 M_3 
\, .  \nonumber
\end{align}

{\bf 3. Dynamics: an example.} Now we are in a position to turn to the flux-tube dynamics. Let us exemplify the inner workings of the formalism on the $\chi_1^2 \chi_7^2$ component 
of the NMHV dodecagon, corresponding at lowest twist to the creation of the scalar $\phi$ at the bottom, its propagation through all intermediate squares and eventual
absorption at the top. The integrand of Eq.\ \re{NMHVcomponents} reads in this case
\begin{align}
\label{ScalarIntegrand}
I^{{\bf\scriptscriptstyle 6} | {\bf\scriptscriptstyle 6} |{\bf\scriptscriptstyle 6} |{\bf\scriptscriptstyle 6} |{\bf\scriptscriptstyle 6} |{\bf\scriptscriptstyle 6} |{\bf\scriptscriptstyle 6} }
(-1, u_1 | \dots | - 1, u_7)
=
\mu_{\phi} (u_1) P_{\phi|\phi} (- u_1 | u_2) \mu_{\phi} (u_2) P_{\phi|\phi} (- u_2 | u_3) \dots  P_{\phi|\phi} (- u_6 | u_7) \mu_{\phi} (u_7)
\, ,
\end{align}
where $\alpha_j = - 1$ in the nomenclature scheme of Ref.\ \cite{Belitsky:2017wdo} according to which $\Phi^{{\bf\scriptscriptstyle 6}}_{-1} = \phi$ with zero helicity 
and unit twist as reminded below. Here $\mu_\phi$ and $P_{\phi | \phi}$ are the scalar measure and its pentagon transition at lowest order in 't Hooft coupling
\cite{Basso:2013aha}. Fourier transform with respect to the rapidities provides the leading OPE contribution to the amplitude
\begin{align}
\label{LOR20000002}
R^{[2,2,2,2,2,2,2]} 
&
= {\rm e}^{- \tau_1 - {\dots} - \tau_7}
\big[
{\rm e}^{\sigma_1+\sigma_2+\sigma_3+\sigma_4+\sigma_5+\sigma_6+\sigma_7}
+
{\rm e}^{ - \sigma_1 + \sigma_2 - \sigma_3 + \sigma_4 - \sigma_5 + \sigma_6 - \sigma_7}
\nonumber\\
&
+
\sum_{j=1}^7 {\rm e}^{\sigma_1+\sigma_2+\sigma_3+\sigma_4+\sigma_5+\sigma_6+\sigma_7 - 2 s_j}
+
\sum_{j=1}^7 \sum_{k=j + 2}^7{\rm e}^{\sigma_1+\sigma_2+\sigma_3+\sigma_4+\sigma_5+\sigma_6+\sigma_7 - 2 s_j - 2 s_k}
\nonumber\\
&\qquad\qquad
+
\sum_{j=1}^7 \sum_{k=j + 2}^7  \sum_{l=k + 2}^7 {\rm e}^{\sigma_1+\sigma_2+\sigma_3+\sigma_4+\sigma_5+\sigma_6+\sigma_7 - 2 s_j - 2 s_k - 2 s_l}
\big]^{-1}
\, ,
\end{align}
which agrees with the corresponding component of the ratio function \re{RinvariantRatio} after the use of the twistors \re{DodecagonTwistors} as can immediately be
verified with the package accompanying Ref.\ \cite{Bourjaily:2013mma}.

The next step is the inclusion of descendants, i.e., adding an arbitrary number of $\psi_{\rm s} \bar{\psi}_{\rm s}$ pairs to the parent scalar. This has to be done in every
intermediate square. Let us begin with just one extra pair at the very bottom, i.e., the process $0 \to \phi \psi_{\rm s} \bar{\psi}_{\rm s} \to \phi \to \dots \to \phi \to 0$. Then, 
the integrand \re{ScalarIntegrand} has to be multiplied by the factor
\begin{align*}
{\rm e}^{- 2 \tau_1}
\int_{C_{\rm s}} \frac{d v_1 \, d v_2}{(2 \pi)^2} 
&
\frac{
\mu_{\psi_{\rm s}} (v_1) \mu_{\psi_{\rm s}} (v_2)
P_{\psi_{\rm s} | \phi} (- v_1 | u_2) P_{\psi_{\rm s} | \phi} (- v_2 | u_2)
}{
|P_{\phi | \psi_{\rm s}} (u_1 | v_1)|^2 |P_{\phi | \psi_{\rm s}} (u_1 | v_2)|^2 |P_{\psi_{\rm s} | \bar\psi_{\rm s}} (v_1 | v_2)|^2
}
\left[ \frac{x [v_1]}{x [v_2]} \right]^{1/2}
\\
&
\times
\frac{1}{6}
[\Pi^{{\bf\scriptscriptstyle 6}}_{0| \phi^{i_1 i_2} \psi^{j_1} \bar\psi_{j_2}}]_{k_1k_2} (0| u_1, v_1, v_2)
[\Pi^{{\bf\scriptscriptstyle 1}}_{\psi^{j_2} \bar\psi_{j_1} \phi_{i_1 i_2} | \phi^{k_1 k_2} }] (- v_2, - v_1, - u_1| u_2)
\, ,
\nonumber
\end{align*}
where the integrations run over the small fermion contours $C_{\rm s}$. The first factor in the above integrand is the factorized form of multiparticle pentagons along with 
the small fermion measures in conventions adopted from Ref.\ \cite{Belitsky:2015efa}. The second factor is the NMHV helicity form factor (on the small sheet) expressed 
via the Zhukowski variable $x[v] \simeq v + O (g^2)$. Last but not least, is the SU(4) tensor part. The latter is quite lengthy but their explicit form can be found in appendices 
to Refs.\ \cite{Belitsky:2017wdo} and \cite{Belitsky:2016vyq} in the order they appear. Substituting the lowest order expressions in 't Hooft coupling in the first line (where 
we already used the fact that the small fermion momentum is of order $g^2$) and evaluating the contour integrals via the Cauchy theorem with the poles arising from the 
matrix part, one finds a very simple result for the factor in question
\begin{align*}
- {\rm e}^{- 2 \tau_1}
(\ft32 + i u_1) (1 + i u_1 + i u_2)
\, .
\end{align*}
This rapidity polynomial can be recast as a differential operator acting on the Fourier exponent in the integrand of Eq.\ \re{NMHVcomponents} and making use of the 
preceding OPE result \re{LOR20000002} successfuly compared with subleading term in the near collinear expansion of  \re{RinvariantRatio}. We have repeated similar
analyses with a $\psi_{\rm s} \bar{\psi}_{\rm s}$ pair in other intermediate squares, i.e., $0 \to \phi \to {\dots} \to \phi \psi_{\rm s} \bar{\psi}_{\rm s} \to {\dots} \to \phi \to 0$ 
and every time found that the integrand acquires a factor
\begin{align*}
&
- {\rm e}^{- 2 \tau_j} (1 + i u_{j - 1} + i u_{j}) (1 + i u_{j} + i u_{j+1})
\, , \quad \mbox{for} \quad
j = 2,3,4,5,6
\, , \\
&
- {\rm e}^{- 2 \tau_7} (1 + i u_6 + i u_7) (\ft32 + i u_7)
\, .
\end{align*}
The procedure was then extended further to up to three pairs either in the same or different squares. We found a recursive pattern which was summarized in the following
proposal for seven-channel conformal block of the flux-tube scalar: 
\begin{align}
&
\mathcal{F}^{[2,2,2,2,2,2,2]}_{1, 1| \dots | 1,1}  (u_1, \tau_1 | \dots | u_7, \tau_7)
\\
&\qquad
=
F_K 
\left. \left(
{
\ft32 + i u_1, 1 + i u_1 + i u_2,  {\dots}, 1 + i u_6 + i u_7, \ft32 + i u_7
\atop
1,1,1,1,1,1,1
} 
\right| 
- {\rm e}^{-2 \tau_1}, {\dots}, - {\rm e}^{-2 \tau_7}
\right)
\, . \nonumber
\end{align}
It is given by the generalization of the Lauricella $F_K$ series, discussed by Saran in Ref.\ \cite{Saran54} for the case of three variables, to $L$ variables
\begin{align}
F_K \!
\left. \left(
{
\alpha_1, \beta_1,  {\dots}, \beta_{L-1}, \alpha_2
\atop
\gamma_1, {\dots}, \gamma_L
} 
\right| 
z_1 , {\dots}, z_L
\right)
=\!\!
\sum_{n_1, \dots, n_L = 0}^\infty
\frac{(\alpha_1)_{n_1} (\beta_1)_{n_1 + n_2} {\dots} (\beta_{L-1})_{n_{L-1} + n_L} (\alpha_2)_{n_L}
}{
(\gamma_1)_{n_1} \dots (\gamma_L)_{n_L}}
\frac{
z_1^{n_1} {\dots} z_L^{n_L}
}{
n_1! \dots n_L! 
}
.
\end{align}
This conjecture was tested numerically to very high orders in the near-collinear expansion against data produced with the help of Ref.\ \cite{Bourjaily:2013mma}
confirming its correctness.

{\bf 4. ${\bit N}$-sided NMHV polygons.} Let us now present a generic form for the flux-tube integrands providing an exact representation for the tree level transitions
\begin{align}
0 \to \Phi^{{\bf\scriptscriptstyle R}}_{s_1, \alpha_1} \to \Phi^{{\bf\scriptscriptstyle R}}_{s_2, \alpha_2} \to {\dots} \to \Phi^{{\bf\scriptscriptstyle R}}_{s_L, \alpha_L} \to 0
\, ,
\end{align}
with the signature $s_j = \pm 1$. Depending on the channel, the parent excitations are
\begin{align}
&
\Phi^{\bf\scriptscriptstyle 6}_{+, 0} = \psi \psi_{\rm s}  \, , \quad \Phi^{\bf\scriptscriptstyle 6}_{+, a > 0} = g_a \psi_{\rm s} \psi_{\rm s}  \, , \\
&
\Phi^{\bf\scriptscriptstyle 6}_{-, - 1} = \phi \, , \quad
\Phi^{\bf\scriptscriptstyle 6}_{-, 0} = \bar\psi \bar\psi_{\rm s} \, , \quad
\Phi^{\bf\scriptscriptstyle 6}_{-, a > 0} = \bar{g}_a \bar\psi_{\rm s}\bar\psi_{\rm s} \, , \nonumber
\end{align}
for the sextet of respective helicities
\begin{align}
h_{+, 0} = 2 \, , \quad h_{+, a} = 2 + a \, , \quad h_{-, -1} = 0 \, , \quad h_{-, 0} = - 1 \, , \quad h_{-, -1} = - 1 - a \, ,
\end{align}
they are
\begin{align}
&
\Phi^{\bf\scriptscriptstyle 4}_{+, 0} = \psi \, , \quad \Phi^{\bf\scriptscriptstyle 4}_{+, a > 0} = g_a \psi_{\rm s} \, , \\
&
\Phi^{\bf\scriptscriptstyle 4}_{-, - 1} = \phi \bar\psi_{\rm s} \, , \quad
\Phi^{\bf\scriptscriptstyle 4}_{-, 0} = \bar\psi \bar\psi_{\rm s}\bar\psi_{\rm s} \, , \quad
\Phi^{\bf\scriptscriptstyle 4}_{-, a > 0} = \bar{g}_a \bar\psi_{\rm s} \bar\psi_{\rm s}\bar\psi_{\rm s} \, , \nonumber
\end{align}
for the quartet of fermionic particles with helicities
\begin{align}
h_{+, 0} = \ft12 \, , \quad h_{+, a} = \ft12 + a \, , \quad h_{-, -1} = - \ft12 \, , \quad h_{-, 0} = - \ft32 \, , \quad h_{-, -1} = - \ft32 - a \, ,
\end{align}
and finally
\begin{align}
&
\Phi^{\bf\scriptscriptstyle 1}_{+, a > 0} = g_a \, , \\ 
&
\Phi^{\bf\scriptscriptstyle 1}_{-, - 2} = \psi \bar\psi_{\rm s} \, , \quad
\Phi^{\bf\scriptscriptstyle 1}_{-, - 1} = \phi \bar\psi_{\rm s} \bar\psi_{\rm s} \, , \quad
\Phi^{\bf\scriptscriptstyle 1}_{-, 0} = \bar\psi \bar\psi_{\rm s} \bar\psi_{\rm s} \bar\psi_{\rm s} \, , \quad
\Phi^{\bf\scriptscriptstyle 1}_{-, a > 0} =  \bar{g}_a \bar\psi_{\rm s} \bar\psi_{\rm s} \bar\psi_{\rm s} \bar\psi_{\rm s}
\, , \nonumber
\end{align}
for singlets with
\begin{align}
h_{+, a} = a \, , \quad h_{-, -2} = 0 \, , \quad h_{-, -1} = -1 \, , \quad h_{-, 0} = - 2 \, , \quad h_{-, -1} = - 2 - a \, ,
\end{align}
The integrands admits the same structure
\begin{align}
&
I^{{\bf\scriptscriptstyle R}| \dots |{\bf\scriptscriptstyle R}}_{s_1| \dots | s_{N-5}} (\alpha_1, u_1 | {\dots} | \alpha_{N-5}, u_{N-5})
=
h^{{\bf\scriptscriptstyle R}| \dots |{\bf\scriptscriptstyle R}}_{s_1| \dots | s_{N-5}} (\alpha_1, u_1 | {\dots} | \alpha_{N-5}, u_{N-5})
\\
&\qquad\qquad
\times
\mu^{{\bf\scriptscriptstyle R}}_{s_1, \alpha_1} (u_1) 
P^{{\bf\scriptscriptstyle R} | {\bf\scriptscriptstyle R}}_{s_1, \alpha_1|s_2, \alpha_2} (- u_1 | u_2) \mu^{{\bf\scriptscriptstyle R}}_{s_2, \alpha_2} (u_2) 
\dots
\nonumber\\
&\qquad\qquad\qquad\qquad
\times \mu^{{\bf\scriptscriptstyle R}}_{s_{L-1}, \alpha_{N - 6}} (u_{N-6}) 
P^{{\bf\scriptscriptstyle R}|{\bf\scriptscriptstyle R}}_{s_{N-6}, \alpha_{N-6}| s_{N-5}, \alpha_{N-5}} (- u_{N-6} | u_{N-5})  
\mu^{{\bf\scriptscriptstyle R}}_{s_{N-5}, \alpha_{N-5}} (u_{N-5})
\, . \nonumber
\end{align}
The helicity NMHV form factors are 
\begin{align}
&
h^{{\bf\scriptscriptstyle 6}| \dots |{\bf\scriptscriptstyle 6}}_{s_1| \dots | s_{N-5}} (\alpha_1, u_1 | {\dots} | \alpha_{N-5}, u_{N-5})
=
(-1)^{1 + \alpha_1}
\left(
\frac{i u_1^{[- \alpha_1/2 - 1]}}{u_1^{[+ \alpha_1/2]}}
\right)^{(1-s_1)/2}
\left(
\frac{i u_{N-5}^{[- \alpha_L/2 - 1]}}{u_{N-5}^{[+ \alpha_{N-5}/2]}}
\right)^{(1+s_{N-5})/2}
\, , \\
&
h^{{\bf\scriptscriptstyle 4}| \dots |{\bf\scriptscriptstyle 4}}_{s_1| \dots | s_L} (\alpha_1, u_1 | {\dots} | \alpha_{N-5}, u_{N-5})
=
(-1)^{1 + \alpha_1}
\left(
\frac{u_1^{[- \alpha_1/2 - 1]} u_1^{[- \alpha_1/2 - 2]}}{i u_1^{[+ \alpha_1/2]}}
\right)^{(1-s_1)/2}
\left(
\frac{1}{i u_{N-5}^{[+ \alpha_{N-5}/2]}}
\right)^{(1+s_{N-5})/2}
\, , \\
&
h^{{\bf\scriptscriptstyle 1}| \dots |{\bf\scriptscriptstyle 1}}_{s_1| \dots | s_{N-5}} (\alpha_1, u_1 | {\dots} | \alpha_{N-5}, u_{N-5})
\nonumber\\
&\qquad\qquad
=
(-1)^{1 + \alpha_1}
\left(
\frac{i u_1^{[- \alpha_1/2 - 1]} u_1^{[- \alpha_1/2 - 2]} u_1^{[- \alpha_1/2 - 3]}}{u_1^{[+ \alpha_1/2]}}
\right)^{(1-s_1)/2}
\left(
\frac{i}{u_{N-5}^{[+ \alpha_{N-5}/2]} u_{N-5}^{[- \alpha_{N-5}/2]}}
\right)^{(1+s_{N-5})/2}
\, , 
\end{align}
where we used the notation $u^{[\alpha]} \equiv u + i \alpha$, while the measure reads
\begin{align}
\mu^{{\bf\scriptscriptstyle R}}_{s, \alpha} (u) = \frac{\Gamma (1 + \ft{\alpha}{2} + i u) \Gamma (1 + \ft{\alpha}{2} - i u)}{\Gamma \big(2 + (2 - r) s + \alpha \big)}
\, ,
\end{align}
and the effective particle pentagon transitions are
\begin{align}
P^{{\bf\scriptscriptstyle R} | {\bf\scriptscriptstyle R}}_{s, \alpha_1|s, \alpha_2} (u_1 | u_2) 
&
=
\frac{
\Gamma (\ft{\alpha_1 - \alpha_2}{2} + i u_1 - i u_2)
\Gamma \big( 2  + (2 - r) s + \ft{\alpha_1 + \alpha_2}{2} - i u_1 + i u_2 \big)
}{
\Gamma (1 + \ft{\alpha_1}{2} + i u_1) \Gamma (1 + \ft{\alpha_1}{2} + i u_2) \Gamma (1 + \ft{\alpha_1 - \alpha_2}{2} - i u_1 + i u_2)
}
\, , \\
P^{{\bf\scriptscriptstyle R} | {\bf\scriptscriptstyle R}}_{s, \alpha_1|- s, \alpha_2} (u_1 | u_2) 
&
=
\frac{(-1)^{\alpha_2}
\Gamma (1 + \ft{\alpha_1 + \alpha_2}{2} + i u_1 - i u_2)
}{
\Gamma (1 + \ft{\alpha_1}{2} + i u_1) \Gamma (1 + \ft{\alpha_1}{2} + i u_2)
}
\, .
\end{align}
The resummation of the infinite number of small fermion-antifermion pairs in all intermediate transitions
\begin{align}
0 \to \Phi^{{\bf\scriptscriptstyle R}}_{s_1, \alpha_1} (\psi \bar\psi_{\rm s})^\infty 
\to 
\Phi^{{\bf\scriptscriptstyle R}}_{s_2, \alpha_2} (\psi \bar\psi_{\rm s})^\infty  
\to {\dots} \to 
\Phi^{{\bf\scriptscriptstyle R}}_{s_L, \alpha_L}  (\psi \bar\psi_{\rm s})^\infty 
\to 0
\, ,
\end{align}
provides the conformal blocks which we sought for
\begin{align}
\label{Multiblock}
\!\!\!\!\!\!\!\!\!\!\mathcal{F}^{[r_1, r_2, \dots, r_{N-5}]}_{h_1, t_1 | h_2, t_2 | \dots | h_{N-5}, t_{N-5}}  (u_1, \tau_1 | u_2, \tau_2 | \dots | u_{N-5}, \tau_{N-5})
\hspace{7cm}
\end{align}
\vspace{-16pt}
\footnotesize
\begin{align*}
=
F_K \!\!
\left. \left(
{\tiny
\frac{|h_1|}{2} \!+\! \frac{2 r_1 + \widehat{r}_1}{4}  \!+\! i u_1,
\!
\frac{|h_1| + |h_2|}{2} \!+\! \frac{\widehat{r}_1 + \widehat{r}_2}{4}  \!+\! i u_1 \!+\! i u_2,  
\dots , 
\frac{|h_{N-5}|}{2} \!+\! \frac{2 r_{N-5} + \widehat{r}_{N-5} - 8}{4} \!+\! i u_{N-5}
\atop
t_1, t_2, \dots, t_{N-5}
}
\right| \!
- {\rm e}^{-2 \tau_1}, {\dots}, - {\rm e}^{-2 \tau_{N-5}}
\!
\right)
\!
, 
\end{align*}
\normalsize
where
\begin{align}
\widehat{r}_j = (4 - r_j) \theta (h_j > 0) + r_j \theta (h_j \leq 0)
\, .
\end{align}
This is the main result of this note.

{\bf 4. Conclusion.} Building up on our previous work dedicated to the heptagon \cite{Belitsky:2017wdo}, we found the multichannel conformal blocks 
\re{Multiblock} for a polygon with any number of sides. The construction was based on resummation over descendants of parent flux-tube excitation
propagating in a given NMHV component of the polygon. The blocks are determined by the generalization of Lauricella hypergeometric series that was 
previously considered by Saran in the particular case of three variables. Multifold integral representation for the latter is available and its extension to the 
generic case should also be looked for since it would be of use for analytical resummation of infinite towers of flux-tube excitations of increasing helicities.

\vspace{0.4cm}

{\bf Acknowledgments.} This research was supported by the U.S. National Science Foundation under the grant PHY-1713125.

%%%%%%%%%%%%%%%%%%%%%%%%%%%%%%%%%%%%%%%%%%%%%%%%%%%%%%%%%%%%%%%%

\vspace{0.4cm}

\appendix

{\bf A. 4-point correlator.} For reader's convenience, let us recall a method for computation of conformal blocks in CFT based on explicit resummation of
descendants, which is adopted in the main body of the paper to the case of scattering amplitudes. Here, it will suffice to discuss the holomorphic sector 
only (or, which is the same, a single light ray) and consider the global sl(2) subalgebra of the Virasoro algebra. Invariance under the sl(2) generators
\begin{align}
\label{Repsl2}
\mathbb{L}^+ = z^2 \partial + 2 d z
\, , \qquad
\mathbb{L}^- = \partial
\, , \qquad
\mathbb{L}^0 = z \partial + d
 \, ,
\end{align}
of the four-point correlator of operators $\mathcal{O}$ of the same conformal dimension $d$,
\begin{align}
\sum_{j = 1}^4 \mathbb{L}^{\pm,0}_j \vev{\mathcal{O} (z_1) \mathcal{O} (z_2) \mathcal{O} (z_3) \mathcal{O} (z_4)} = 0
\, ,
\end{align}
fixes its form 
\begin{align}
\vev{\mathcal{O} (z_1) \mathcal{O} (z_2) \mathcal{O} (z_3) \mathcal{O} (z_4)}
= 
\frac{\mathcal{F}_4 (w)}{z_{13}^{2 d} z_{24}^{2 d}}
\, , \qquad
w = \frac{z_{12} z_{34}}{z_{13} z_{24}}
\, .
\end{align}
up to a function of the conformal cross ratio $w$. Let us choose a conformal frame, by setting
\begin{align}
z_4 = 0 \, , \qquad z_2 = 1 \, , \qquad z_1 = \infty.
\end{align}
The operator--state correspondence (in the radial quantization) allows us to write
\begin{align}
\lim_{z_1 \to \infty} z_1^{2d}\vev{\mathcal{O} (z_1) \mathcal{O} (1) \mathcal{O} (z_3) \mathcal{O} (0)} = \bra{d}  \mathcal{O} (1) \mathcal{O} (z_3) \ket{d} 
\, .
\end{align}

To compute conformal blocks let us assume that the intermediate state is a primary state $\ket{\Delta}$, i.e., $L^+ \ket{\Delta} = 0$, of dimension $\Delta$ and its 
descendants are
\begin{align}
\ket{\Delta,k} \equiv (L^-)^k \ket{\Delta}
\, , k > 0
\, .
\end{align}
Obviously, $\bra{\Delta,k} \equiv \bra{\Delta} (L^+)^k$. Here $L^{\pm,0}$ are operators acting on the Hilbert space of states with the representation \re{Repsl2} on the primary 
fields $\Phi_\Delta$. We can project on these with 
\begin{align}
\Pi_\Delta = \sum_{k = 0}^\infty \frac{\ket{\Delta,k} \bra{\Delta, k}}{N_{\Delta,k}} 
\, , 
\end{align}
obeying $\Pi_\Delta^2 = \Pi_\Delta$, with the normalization $N_{\Delta, k}  = \vev{\Delta, k | \Delta,k}$. Such that
\begin{align}
\label{F4}
\mathcal{F}_4 (z_3) 
= \bra{d}  \mathcal{O} (1)\Pi_\Delta  \mathcal{O} (z_3) \ket{d} 
= \sum_{k = 0}^\infty \frac{1}{N_{\Delta, k}}  \bra{d} \mathcal{O} (1) (L^-)^k \ket{\Delta} \bra{\Delta} (L^+)^k \mathcal{O} (z_3) \ket{d} 
\, . 
\end{align}
The calculation of the matrix elements involved is straightforward making use of the sl(2) algebra. The normalization prefactor reads
\begin{align}
N_{\Delta, k}  = \vev{\Delta | [(L^+)^k , (L^-)^k ] |\Delta} 
= k! (2 \Delta)_k
\, ,
\end{align}
which is a generalization of the elementary commutation relation
\begin{align*}
\vev{\Delta | [(L^+)^2 , (L^-)^2 ] |\Delta} = 2 (2\Delta + 1) \vev{\Delta | [L^+ , L^- ] |\Delta} = 2! \, 2 \Delta (2\Delta + 1) \vev{\Delta|\Delta}
\, .
\end{align*}
The matrix element in the numerator of the right-hand side of Eq.\ \re{F4} reads
\begin{align}
\bra{\Delta} (L^+)^k  \mathcal{O} (z_3) \ket{d} 
=
\bra{\Delta} [L^+, [L^+, \dots [L^+,  \mathcal{O} (z_3)] \dots]] \ket{d} 
=
(\mathbb{L}^+) ^k \bra{\Delta}  \mathcal{O} (z_3) \ket{d} 
\, .
\end{align}
Since
\begin{align}
\bra{\Delta}  \mathcal{O} (z_3) \ket{d}  = 1/z_3^{2d - \Delta}
\, ,
\end{align}
is just the three-point function (fixed up to an overall normalization (set here to one) by conformal symmetry), we immediately find after repetitive differentiation
\begin{align}
\bra{\Delta} (L^+)^k \mathcal{O} (z_3) \ket{d} = (\Delta)_k/z_3^{2d - \Delta - k}
\, .
\end{align}
Putting everything together, we find for $\mathcal{F}_4 (z_3)$
\begin{align}
\mathcal{F}_4 (z_3) = z_3^{\Delta - 2j}\sum_{k = 0}^\infty \frac{(\Delta)_k^2 }{k! (2\Delta)_k} z^k
=
z_3^{\Delta - 2d} {_2 F_1} \left. \left( { \Delta, \Delta \atop 2\Delta }\right| z_3 \right)
\, , 
\end{align}
which is a well-known result \cite{Ferrara:1973eg}.

The same result can be obtained making use of the eigenvalue equation for the quadratic Casimir of the sl(2) algebra,
\begin{align}
\mathbb{C}_2 = \ft{1}{2} \left( \mathbb{L}^+ \mathbb{L}^- + \mathbb{L}^- \mathbb{L}^+ \right) - (\mathbb{L}^0)^2
\end{align}
in a given OPE channel. For instance, in the (34)-channel, which is the same as the (12)-channel,
\begin{align}
\mathbb{L}_{34}^{\pm, 0} = \mathbb{L}_{3}^{\pm, 0} + \mathbb{L}_{4}^{\pm, 0}
\, ,
\end{align}
the equation 
\begin{align}
\mathbb{C}_{2, (34)} \vev{\mathcal{O} (z_1) \mathcal{O} (z_2) \mathcal{O} (z_3) \mathcal{O} (z_4)}
= 
\Delta (1 - \Delta) \vev{\mathcal{O} (z_1) \mathcal{O} (z_2) \mathcal{O} (z_3) \mathcal{O} (z_4)}
\, ,
\end{align}
immediately implies that $\mathcal{F}_4 (w)$ obeys
\begin{align}
w^2 (w - 1) \mathcal{F}_4^{\prime\prime} (w) + \left[ 4 d w (w-1) - w^2 \right] \mathcal{F}_4^{\prime} (w) 
+
\left[ 2 d (1 + 2d (w - 1))+ \Delta (\Delta - 1) \right] \mathcal{F}_4 (w)
= 0
\, .
\end{align}
It has two solutions
\begin{align}
\mathcal{F}_4 (w)
= 
w^{\Delta - 2d} {_2 F_1} \left. \left( { \Delta, \Delta \atop 2\Delta }\right| w \right)
+
{\rm c\,}
w^{1 - \Delta - 2d} {_2 F_1} \left. \left( { 1 - \Delta, 1 - \Delta \atop 2 - 2 \Delta }\right| w \right)
\, .
\end{align}
However, the second one does not possess correct asymptotic behavior and thus have to be dropped, i.e., ${\rm c} = 0$. This way, we recover our
previous result for the conformal block.

{\bf B. Conformal frame for polygons.} The choices made in the body for the elements of the symmetry matrices of middles squares in the tessellation of a 
generic polygon correspond to the following conformal cross ratios \cite{Alday:2010vh}
\begin{align}
{\rm e}^{\tau_{2j+1}} 
&= 
\frac{(-j-1,j+1,j+2,j+3)(-j-2,-j-1,-j,j+2)}{(-j-2,-j-1,j+2,j+3)(-j-1,-j,j+1,j+2)}
\, , \\
{\rm e}^{\tau_{2j+1} + \sigma_{2j+1} - i \phi_{2j+1}} 
&= 
\frac{(-j-2,-j-1,-j,-j+1)(-j-1,-j,j+2,j+3)}{(-j-2,-j-1,-j,j+3)(-j-1,-j,-j+1,j+2)}
\, , \\
{\rm e}^{\tau_{2j+1} + \sigma_{2j+1} + i \phi_{2j+1}} 
&= 
\frac{(j+1,j+2,j+3,j+4)(-j-1,-j,j+2,j+3)}{(-j-1,j+2,j+3,j+4)(-j,j+1,j+2,j+3)}
\, , \\
{\rm e}^{\tau_{2j}} 
&= 
\frac{(-j,j+1,j+2,j+3)(-j-1,-j,-j+1,j+2)}{(-j-1,-j,j+2,j+3)(-j,-j+1,j+1,j+2)}
\, , \\
{\rm e}^{\tau_{2j} + \sigma_{2j} - i \phi_{2j}}
&= 
\frac{(-j-1,-j,j+1,j+2)(j,j+1,j+2,j+3)}{(-j-1,j+1,j+2,j+3)(-j,j,j+1,j+2)}
\, , \\
{\rm e}^{\tau_{2j} + \sigma_{2j} + i \phi_{2j}} 
&= 
\frac{(-j-2,-j-1,-j,-j+1)(-j-1,-j,j+1,j+2)}{(-j-2,-j-1,-j,j+2)(-j-1,-j,-j+1,j+1)}
\, .
\end{align}
Here the odd and even ratios have different form due to opposite orientation of overlapping sequential pentagons.

%%%%%%  Bibliography %%%%%%%%%%%%%%%%%%%%%%%%%%%%%%%%%%%%%%%%%%%%


\begin{thebibliography}{100}
%
\bibitem{Basso:2013vsa}  
B.~Basso, A.~Sever, P.~Vieira,
``Spacetime and flux tube S-matrices at finite coupling for N=4 supersymmetric Yang-Mills theory,''
Phys.\ Rev.\ Lett.\  {\bf 111} (2013) 091602
[arXiv:1303.1396 [hep-th]].
%%CITATION = ARXIV:1303.1396;%%
%
\bibitem{Alday:2010ku}
L.F.~Alday, D.~Gaiotto, J.~Maldacena, A.~Sever, P.~Vieira,
``An Operator Product Expansion for Polygonal null Wilson Loops,''
JHEP {\bf 1104} (2011) 088 [arXiv:1006.2788 [hep-th]].
%%CITATION = doi:10.1007/JHEP04(2011)088;%%
%
\bibitem{Alday:2007hr}  
L.F.~Alday, J.M.~Maldacena,
``Gluon scattering amplitudes at strong coupling,''
JHEP {\bf 0706} (2007) 064 [arXiv:0705.0303 [hep-th]].
%%CITATION = ARXIV:0705.0303;%%
%
\bibitem{Drummond:2007cf}
J.M.~Drummond, J.~Henn, G.P.~Korchemsky, E.~Sokatchev,
``On planar gluon amplitudes/Wilson loops duality,''
Nucl.\ Phys.\ B {\bf 795} (2008) 52 [arXiv:0709.2368 [hep-th]].
%%CITATION = ARXIV:0709.2368;%%
%
\bibitem{Brandhuber:2007yx}
A.~Brandhuber, P.~Heslop, G.~Travaglini,
``MHV amplitudes in N=4 super Yang-Mills and Wilson loops,''
Nucl.\ Phys.\ B {\bf 794} (2008) 231 [arXiv:0707.1153 [hep-th]].
%%CITATION = ARXIV:0707.1153;%%
%
\bibitem{CaronHuot:2010ek}
S.~Caron-Huot,
``Notes on the scattering amplitude / Wilson loop duality,''
JHEP {\bf 1107} (2011) 058 [arXiv:1010.1167 [hep-th]].
%%CITATION = ARXIV:1010.1167;%%
%
\bibitem{Mason:2010yk}
L.J.~Mason, D.~Skinner,
``The Complete Planar S-matrix of N=4 SYM as a Wilson Loop in Twistor Space,''
JHEP {\bf 1012} (2010) 018 [arXiv:1009.2225 [hep-th]].
%%CITATION = ARXIV:1009.2225;%%
%
\bibitem{Belitsky:2011zm}
A.V.~Belitsky, G.P.~Korchemsky, E.~Sokatchev,
``Are scattering amplitudes dual to super Wilson loops?,''
Nucl.\ Phys.\ B {\bf 855} (2012) 333 [arXiv:1103.3008 [hep-th]].
%%CITATION = ARXIV:1103.3008;%%
%
\bibitem{Gaiotto:2010fk} 
D.~Gaiotto, J.~Maldacena, A.~Sever, P.~Vieira,
``Bootstrapping Null Polygon Wilson Loops,''
JHEP {\bf 1103} (2011) 092 [arXiv:1010.5009 [hep-th]].
%%CITATION = doi:10.1007/JHEP03(2011)092;%%
%
\bibitem{Basso:2010in}
B.~Basso,
``Exciting the GKP string at any coupling,''
Nucl.\ Phys.\ B {\bf 857} (2012) 254 [arXiv:1010.5237 [hep-th]].
%%CITATION = doi:10.1016/j.nuclphysb.2011.12.010;%%
%
\bibitem{Gaiotto:2011dt}
D.~Gaiotto, J.~Maldacena, A.~Sever, P.~Vieira,
``Pulling the straps of polygons,''
JHEP {\bf 1112}, 011 (2011) [arXiv:1102.0062 [hep-th]].
%%CITATION = doi:10.1007/JHEP12(2011)011;%%
%
\bibitem{Cordova:2016woh}
L.~Cordova,
``Hexagon POPE: effective particles and tree level resummation,''
JHEP {\bf 1701} (2017) 051 [arXiv:1606.00423 [hep-th]].
%%CITATION = doi:10.1007/JHEP01(2017)051;%%
%
\bibitem{Sever:2011pc}
A.~Sever, P.~Vieira,
``Multichannel Conformal Blocks for Polygon Wilson Loops,''
JHEP {\bf 1201} (2012) 070  [arXiv:1105.5748 [hep-th]].
%%CITATION = doi:10.1007/JHEP01(2012)070;%%
%
\bibitem{Belitsky:2017wdo}
A.V.~Belitsky,
``Resummed tree heptagon,''
arXiv:1710.06567 [hep-th].
%%CITATION = ARXIV:1710.06567;%%
%
\bibitem{Drummond:2008vq}
J.M.~Drummond, J.~Henn, G.P.~Korchemsky, E.~Sokatchev,
``Dual superconformal symmetry of scattering amplitudes in N=4 super-Yang-Mills theory,''
Nucl.\ Phys.\ B {\bf 828} (2010) 317 [arXiv:0807.1095 [hep-th]].
%%CITATION = doi:10.1016/j.nuclphysb.2009.11.022;%%
%
\bibitem{Mason:2009qx}
L.J.~Mason, D.~Skinner,
``Dual Superconformal Invariance, Momentum Twistors and Grassmannians,''
JHEP {\bf 0911} (2009) 045 [arXiv:0909.0250 [hep-th]].
%%CITATION = doi:10.1088/1126-6708/2009/11/045;%%
%
\bibitem{Basso:2013aha}  
B.~Basso, A.~Sever, P.~Vieira,
``Space-time S-matrix and flux tube S-matrix II. Extracting and matching data,''
JHEP {\bf 1401} (2014) 008 [arXiv:1306.2058 [hep-th]].
%%CITATION = ARXIV:1306.2058;%%
%
\bibitem{Bourjaily:2013mma}
J.L.~Bourjaily, S.~Caron-Huot, J.~Trnka,
``Dual-Conformal Regularization of Infrared Loop Divergences and the Chiral Box Expansion,''
JHEP {\bf 1501} (2015) 001 [arXiv:1303.4734 [hep-th]].
%%CITATION = doi:10.1007/JHEP01(2015)001;%%
%
\bibitem{Belitsky:2015efa}
A.V.~Belitsky,
``On factorization of multiparticle pentagons,''
Nucl.\ Phys.\ B {\bf 897} (2015) 346 [arXiv:1501.06860 [hep-th]].
%%CITATION = ARXIV:1501.06860;%%
%
\bibitem{Belitsky:2016vyq}
 A.V.~Belitsky,
``Matrix pentagons,''
Nucl.\ Phys.\ B {\bf 923}, 588 (2017)  [arXiv:1607.06555 [hep-th]].
%%CITATION = doi:10.1016/j.nuclphysb.2017.08.011;%%
%
\bibitem{Saran54}
S. Saran,
``Hypergeometric functions of three variables,"
Ganita {\bf 5} (1954) 71;
%
``Integrals associated with hypergeometric functions of three variables,"
Proceedings of the Indian National Science Academy {\bf 21A} (1955) 83.
%
\bibitem{Alday:2010vh}
L.F.~Alday, J.~Maldacena, A.~Sever, P.~Vieira,
``Y-system for Scattering Amplitudes,''
J.\ Phys.\ A {\bf 43} (2010) 485401 [arXiv:1002.2459 [hep-th]].
%%CITATION = doi:10.1088/1751-8113/43/48/485401;%%
%  
\bibitem{Basso:2014koa} 
B.~Basso, A.~Sever, P.~Vieira,
``Space-time S-matrix and flux-tube S-matrix III. The two-particle contributions,''
JHEP {\bf 1408} (2014) 085 [arXiv:1402.3307 [hep-th]].
%%CITATION = ARXIV:1402.3307;%%
%
\bibitem{Belitsky:2014sla} 
A.V.~Belitsky,
``Nonsinglet pentagons and NHMV amplitudes,''
Nucl.\ Phys.\ B {\bf 896} (2015) 493
[arXiv:1407.2853 [hep-th]].
%%CITATION = ARXIV:1407.2853;%%
%
\bibitem{Basso:2014nra}
B.~Basso, A.~Sever, P.~Vieira,
``Space-time S-matrix and flux-tube S-matrix IV. Gluons and fusion,''
JHEP {\bf 1409} (2014) 149 [arXiv:1407.1736 [hep-th]].
%%CITATION = ARXIV:1407.1736;%%
%
\bibitem{Belitsky:2014lta} 
A.V.~Belitsky,
``Fermionic pentagons and NMHV hexagon,''
Nucl.\ Phys.\ B {\bf 894} (2015) 108 [arXiv:1410.2534 [hep-th]].
%%CITATION = ARXIV:1410.2534;%%
%
\bibitem{Basso:2014hfa}
B.~Basso, J.~Caetano, L.~Cordova, A.~Sever, P.~Vieira,
``OPE for all helicity amplitudes,''
JHEP {\bf 1508} (2015) 018  [arXiv:1412.1132 [hep-th]].
%%CITATION = doi:10.1007/JHEP08(2015)018;%%
%
\bibitem{Basso:2015rta}    
B.~Basso, J.~Caetano, L.~Cordova, A.~Sever, P.~Vieira,
``OPE for all helicity amplitudes II. Form factors and data analysis,''
JHEP {\bf 1512} (2015) 088  [arXiv:1508.02987 [hep-th]].
%%CITATION = doi:10.1007/JHEP12(2015)088;%%
%
\bibitem{Ferrara:1973eg}
S.~Ferrara, R.~Gatto, A.F.~Grillo,
``Conformal algebra in space-time and operator product expansion,''
Springer Tracts Mod.\ Phys.\  {\bf 67} (1973) 1.
 %%CITATION = doi:10.1007/BFb0111104;%%
%
\end{thebibliography}
\end{document}